\documentclass[useAMS,usenatbib]{mn2e}
\newcommand{\hdos}{H$_2$}
\newcommand{\sco}{$S_\text{CO}\Delta v$}
\newcommand{\msun}{M$_{\odot}$}

\usepackage{graphicx}
\usepackage{natbib}
\usepackage{amssymb}
\usepackage{amsmath}
\usepackage{lscape}
\usepackage{longtable}
\usepackage{color}
\title[ALMA Observations of YMCs in the Antennae]{Constraining globular cluster formation through studies of young massive clusters - V. ALMA observations of clusters in the Antennae}
\author[I. Cabrera-Ziri et al.]{I. Cabrera-Ziri$^{1,2}$\thanks{ICZ: I.CabreraZiriCastro@2013.ljmu.ac.uk},   N. Bastian$^{1}$, S. N. Longmore$^{1}$, C. Brogan$^{3}$, K. Hollyhead$^{1}$,
\and S. S. Larsen$^{4}$, B. Whitmore$^{5}$, K. Johnson$^6$, R. Chandar$^{7}$, J. D. Henshaw$^{1}$,
\and B. Davies$^{1}$, J. E. Hibbard$^{3}$\\
$^{1}$ Astrophysics Research Institute, Liverpool John Moores University, 146 Brownlow Hill, Liverpool L3 5RF, UK\\
$^{2}$ European Southern Observatory, Karl-Schwarzschild-Stra{\ss}e 2, D-85748 Garching bei M{\"u}nchen, Germany\\
$^{3}$  National Radio Astronomy Observatory, 520 Edgemont Rd, Charlottesville, VA 22903, USA\\
$^{4}$ Department of Astrophysics/IMAPP, Radboud University Nijmegen, PO Box 9010, NL-6500 GL Nijmegen, the Netherlands\\
$^{5}$ Space Telescope Science Institute, 3700 San Martin Drive, Baltimore, MD 21218, USA\\
$^{6}$ Department of Astronomy, University of Virginia, P.O. Box 400325, Charlottesville, VA 22904-4325, USA\\
$^{7}$ Department of Physics \& Astronomy, University of Toledo, Toledo, OH 43606, USA\\
}
\begin{document}

\date{Accepted XXX. Received XXX; in original form XXX}

\pagerange{\pageref{firstpage}--\pageref{lastpage}} \pubyear{2014}

\maketitle

\label{firstpage}

\begin{abstract}

Some formation scenarios that have been put forward to explain multiple populations within Globular Clusters (GCs) require that the young massive cluster have large reservoirs of cold gas within them, which is necessary to form future generations of stars. In this paper we use deep observations taken with Atacama Large Millimeter/sub-millimeter Array (ALMA) to assess the amount of molecular gas within 3 young ($50-200$ Myr) massive ($\sim10^6$ M$_\odot$) clusters in the Antennae galaxies. No significant CO(3--2) emission was found associated with any of the three clusters. We place upper limits for the molecular gas within these clusters of $\sim1\times10^5$ M$_\odot$ (or $<9$\% of the current stellar mass). We briefly review different scenarios that propose multiple episodes of star formation and discuss some of their assumptions and implications. Our results are in tension with the predictions of GC formation scenarios that expect large reservoirs of cool gas within young massive clusters at these ages.

\end{abstract}


\section{Introduction}
\label{sec:intro}

Since the discovery of the evidence of multiple stellar populations in the colour--magnitude diagrams (CMD) and chemical abundances of globular clusters (GCs), the classic view of GCs as simple stellar populations has been called into question (cf. \citealt{Gratton:2012p2005}). Different scenarios have been put forward to account for these ``anomalous'' observed features. 

The majority of the proposed formation scenarios for GCs adopt multiple generations of stars in order to explain the observed anomalies. Essentially the main concept is that the chemically processed ejecta of some kinds of stars from the first generation (\emph{polluter} stars) breed a second generation of stars with the observed ``anomalies''. Several kinds of stars have been suggested to be the \emph{polluters} namely: Asymptotic Giant Branch (AGB) stars (e.g. \citealt{Dercole:2008p2154,Renzini:2008p2648,Conroy:2011p1997} - hereafter D08, R08 and CS11 respectively), fast rotating massive stars, FRMS (e.g. \citealt{Decressin:2009p2155,Krause:2013p2563}), and massive stars in interacting binary systems \citep{DeMink:2009p2156}.

In all of these scenarios, the ejecta from the \emph{polluters} is expected to cool and sink into 
 the gravitational potential well of the cluster where it will spawn the future generations of stars. 
These models all require additional ``pristine" material (i.e. the gas from which the first generation was born) to be (re)accreted by the clusters from their surroundings in order to match the chemical patterns found in GCs, or alternatively that some left over gas from the first generation stays bound in the centre of the cluster and mixes with the polluted material and subsequently forms stars (e.g. D08; CS11; \citealt{Krause:2013p2563}). 

Each type of \emph{polluter} star is expected to have a different timescale for the second episode of star formation. For example, the AGB scenario predicts that the second star formation episode begins $\sim30$~Myr after the birth of the first generation (when the super-AGB stars begin to eject their low velocity winds). This then continues until $\sim80-100$~Myr, when it is truncated by prompt SNe Ia explosions. For this model this is necessary in order to match the observed abundance trends in GCs (e.g. D08 -  as lower mass AGBs stars, with longer lifetimes, have different abundance yields).  However, other authors have suggested that this maximum needs be pushed to later times ($\sim200-300$~Myr) in order to allow the Lyman-Werner flux density to drop sufficiently for the gas to cool and form new stars (CS11, see also R08 for an alternative long timescale AGB scenario - few $10^8$ yr). On the other hand, the FRMS and the interacting binaries scenarios work on much shorter timescales (a few Myr) for the formation of the second generation.

In this series of papers, we have used observations of young massive clusters (YMCs) often seen as young GCs (e.g. \citealt{Schweizer:1998p2017}), to probe the different models for GC formation. Some of these studies were focused on the properties of the gas within YMCs. For example \cite{Bastian:2013p2022} analysed over 100 YMCs for evidence of ionised gas ([O{\sc iii}] or H$\beta$) due to ongoing star formation, and found none.  The authors set an upper limit of 1--2\% of the cluster stellar mass to be currently forming within the cluster.
In a following work, \citet{Bastian:2014} placed limits on the amount of H{\sc i} and/or dust found within 13 LMC/SMC YMCs (with ages between 15 and 300 Myr, and masses between $10^4$ and $2\times 10^5$ M$_\odot$) to be $< 1\%$ of the stellar mass, this result is at odds with the expectations of some scenarios (e.g. CS11). In addition, \cite{Longmore2015} has shown that high levels of extinction/dust would be expected for the clusters if there were to be such large reservoirs of gas/dust within these clusters as suggested by D08.  \citet{Bastian:2014b}  also used YMCs to test a prediction from the FRMS scenario, namely that massive young clusters should remain embedded in the natal gas cloud for the first $20-30$~Myr of the cluster's life.  However, it was found that YMCs, independent of their mass, are very efficient at expelling any gas within/near them, beginning $1-3$~Myr after their formation. Similar results have been found in YMCs in the Antennae, where they appear to have removed the molecular gas within them over the first 10 Myr of their life clearing the gas out to a radius of $\sim$200 pc \citep{Whitmore:2014p2682}. 


In the current paper we look for evidence of molecular gas necessary to form the second generation of stars within young GCs with Atacama Large Millimeter/sub-millimeter Array (ALMA) observations. Our study is centred on 3 young ($<200$ Myr) massive ($\sim10^6$ M$_\odot$) clusters in the Antennae galaxies (W32187, W32604 and W31730). We estimate the escape velocities for these clusters to be between 50--130 km/s, well above the velocity of the stellar ejecta of the different \emph{polluters}, allowing them in principle to be able to retain the ejecta from these stars. These clusters have very little reddening, with $E(B-V)$ ranging from 0.06 to 0.5 mag, and are amongst the 12 brightest/most massive clusters in the Antennae galaxies, which makes them ideal for this kind of study (\citealt{Whitmore:2010p2561}, hereafter W10).

The paper is organised as follows:  In \S \ref{sec:data} we present the ALMA data of these clusters and in \S \ref{sec:masses} we show the procedure used to estimate the H$_2$ masses for these clusters.  We discuss our results and present our conclusions in \S \ref{sec:discussion} and \S\ref{sec:conclusions}, respectively.

\section{Data}
\label{sec:data}

We made use of ``Band 7" (345 GHz) ALMA observations of the Antennae galaxies (NGC 4038/39) from Cycle 0 project 2011.0.00876. A detailed description of these observations and data reduction can be found in \cite{Whitmore:2014p2682}, along with a formal introduction to the highest spatial resolution CO (3--2) observations of the overlap region in the Antennae.

\begin{figure}
\includegraphics[width= 80mm,height=80mm]{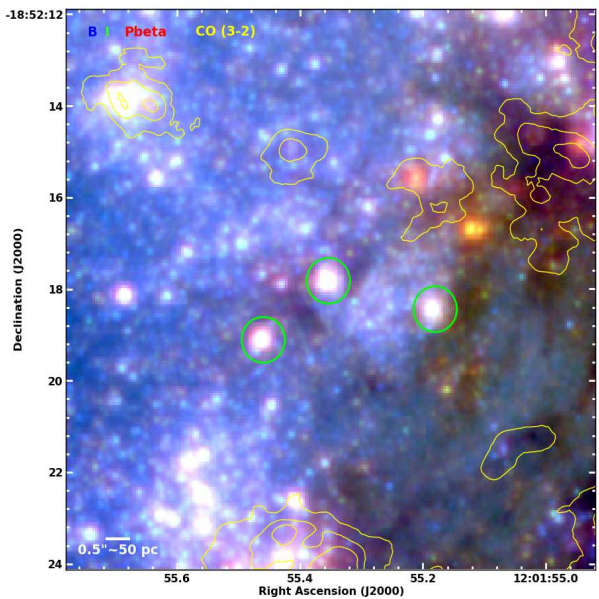}
\caption{Zoom to the region containing the clusters. This composite image was made with $HST$ observations where the red, green and blue colours are from Pa$\beta$, $I$-band, $B$-band images (from \citealt{Whitmore:2014p2682}). The contours overlaid in yellow are CO (3--2), representing flux levels of 4, 24, and 48 K km/s. The green circles highlight the clusters of interest.}
\label{image}
\end{figure}

In Fig. \ref{image} we present a colour composite image of the clusters of our analysis (based on HST imaging); W32187, W32604 and W31730 (from east to west respectively, W10)\footnote{ a.k.a. cluster S2\_1, [W99]15 and S1\_2 in \cite{Mengel:2008p2564}.}. The clusters of interest are highlighted with green circles. The contours indicate the integrated CO emission, from which it can already been seen that no CO emission is associated with the clusters.

Only these clusters fulfilled the following selection criteria 1) be located in on our ALMA field of view and away of any background CO emission 2) were young ($<200$ Myr) and massive ($>10^6$ M$_\odot$) 3) have near-IR spectroscopy used to determine radial velocities and velocity dispersions. There are 4 more clusters that satisfy criteria 1) and 2) but they lack near-infrared spectroscopy (i.e. with no radial velocities nor velocity dispersions), these clusters will be studied in a future work.

\begin{figure*}
\includegraphics[width= 180mm,height=84mm]{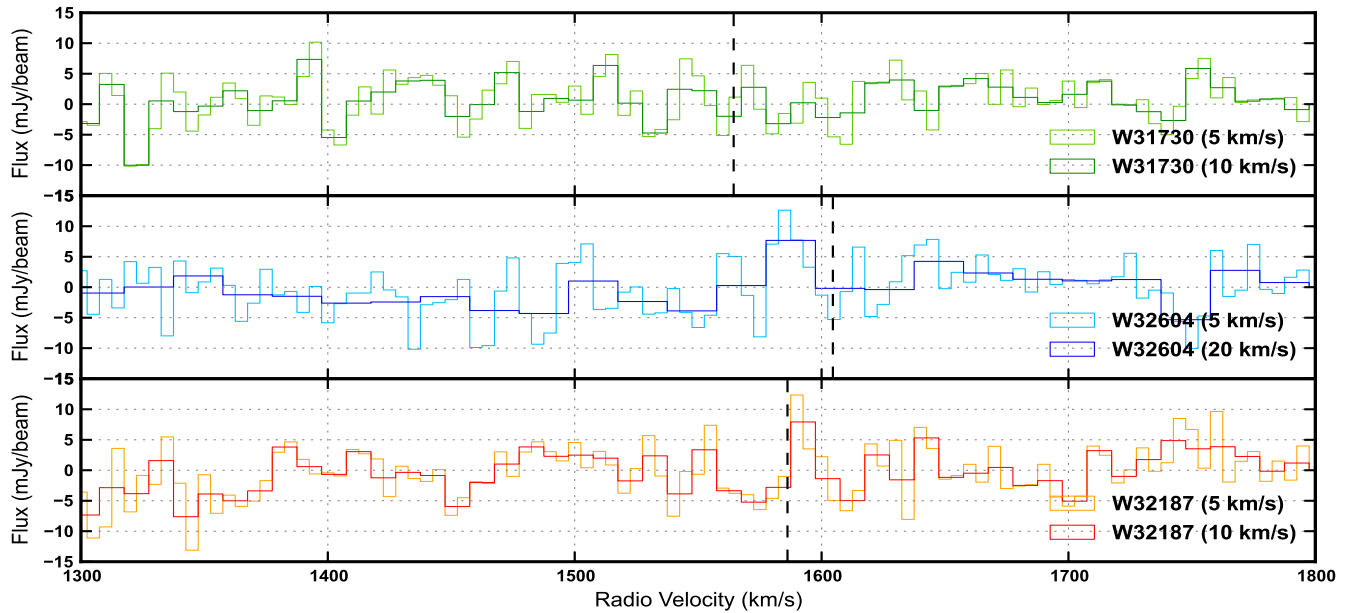}
\caption{ALMA spectra of the 3 young massive clusters in the Antennae galaxies. The spectrum of each cluster is presented with the highest resolution (bins of 5 km/s) and also binned to the approximate velocity dispersion reported in M08. There is no detection of any significant CO (3--2) emission in any of the clusters studied in this work. The vertical dashed lines represents the radial velocities of each cluster from M08 corrected to LSRK. The position of the extracted spectra and rms values of the flux are reported in Table \ref{table1}.}
\label{spec}
\end{figure*}

Fig. \ref{spec} shows the spectra of the (primary beam corrected) CO (3--2) emission towards the three clusters in units of mJy/beam. For these observations the channel spacing is 5 km/s and the synthesised beam size is 0.56"$\times$0.43". All spectra were extracted from single beams taken from the approximate I-band intensity peak. These positions are given in Table \ref{table1} together with the characteristics of each cluster (see \S\ref{sec:masses}). We noted a systematic offset of -1.088" and 1.65" in R.A. and Dec. respectively, in the positions of the clusters reported in table 7 in W10 with respect to the ones in the HST images. 


\section{Molecular gas and stellar mass estimates}
\label{sec:masses}
%

Following equation (3) in \cite{Bolatto:2013p2560}:

\begin{equation*}
M_{\mbox{mol}} = 1.05\times 10^4 \left( \frac{X_\text{CO}}{2 \times 10^{20} \text{ cm}^{-2} (\text{K km/s})^{-1}}\right) \frac{S_\text{CO}\Delta vD_L^2}{(1+z)}\mbox{ M}_\odot
\label{eq:Mmol}
\end{equation*}

\noindent we derived the H$_2$ gas masses for these clusters. For the H$_2$ mass estimates we assume a CO (1--0)-to-\hdos~conversion factor of $ X_\text{CO} = 2 \times 10^{20} \text{ cm}^{-2} (\text{K km/s})^{-1}$, a luminosity distance of $D_\text{L}= 21.8$ Mpc to the Antennae galaxies and a redshift of $z=0.005688$. Finally \sco, is the integrated CO (1--0) line flux density, in our case given by

\begin{equation*}
S_\text{CO}\Delta v = \frac{F_\text{rms} \Delta v}{r_{31}}
\label{eq:sco}
\end{equation*}

\noindent where $F_\text{rms}$ is the rms value of the flux of our ALMA CO (3--2) spectra shown in Fig. \ref{spec} (we assumed this value to be the standard deviation of the flux, $\sigma$, for these spectra); $\Delta v$ is the channel spacing of the spectra (5 km/s for the high resolution spectra and 10 or 20 km/s for the low resolution ones, see Fig. \ref{spec}); and $r_{31}$ is the CO (3--2) to CO (1--0) line ratio. We adopted $r_{31}=0.792$, this value was obtained averaging the measurements of $r_{31}$ reported for the Antennae on Table 2 from \cite{Zhu:2003p2569}.

In Table \ref{table1} we present the characteristics of these clusters. The ages and photometric masses are from W10. We remeasured the sizes of these clusters and used these values to recalculate the dynamical masses of \cite{Mengel:2008p2564}, hereafter M08. For the sizes, we made an empirical PSF based on ACS $V$-band drizzled images and used ISHAPE \citep{Larsen:1999p2562} to measure the effective radii adopting different profiles (see \citealt{Bastian:2009p1988} for more details on the technique). The derived sizes were 2.7, 1.7, and 2 times larger than reported in M08 for W32187, W32604 and W31730 respectively, which corresponds to an increase of the dynamical mass of the same proportion. We find excellent agreement between the photometric and dynamical masses.

In this table we also include $F_\text{rms}$, \sco~and \hdos ~mass. We note that the measurements presented here are $1\sigma$ upper limits for \sco~and \hdos ~mass. Already with this data we do not find clear evidence of molecular emission above the $1\sigma$ level in any of these spectra.




\begin{table*}
\caption{Clusters' ages, spectra position, photometric masses, projected half-light radii, dynamical masses, flux spectra rms, \sco~and H$_2$ mass. Values in parenthesis correspond to the binned spectra.}
\begin{center}
\begin{tabular}{cccccccccc}
Cluster & $\alpha$ & $\delta$ & log(Age)$^a$ & M$_\text{ph}$$^a$ & r$_\text{hp}$ & M$_{dyn}$ & $F_\text{rms}$ & \sco & \hdos~mass \\
ID & & &  & ($\times10^6$ \msun) & (pc) &($\times10^6$ \msun) & (mJy/beam) &  (mJy km/s) & ($\times10^5$ \msun) \\
\hline W31730 & 12 01 55.462 & -18 52 19.108 & 7.7  & 1.7 & 4.1 &1.4 & 3.98 (3.21)$^b$ & 25.11 (40.58)$^b$ &1.25 (2.01)$^b$\\
W32604 & 12 01 55.354 & -18 52 17.824 & 7.7 & 1.6 & 3.4 &2.4 & 4.48 (2.84)$^c$ &  28.29 (71.73)$^c$ &1.40 (3.56)$^c$\\
W32187 & 12 01 55.184 & -18 52 18.442 & 8.3 & 1.9 & 8.1 &1.8 & 4.43 (3.47)$^b$ & 27.98 (43.84)$^b$ &1.38 (2.17)$^b$\\
\end{tabular}
\end{center}
$^a$ from W10. We assumed conservative errors for these age estimates of $\pm50\%$.\raggedright \\
$^b$ value for the spectrum binned to 10 km/s (similar to the velocity dispersion of 11.5 km/s reported in M08 for W31730 and W32187).\raggedright \\
$^c$ value for the spectrum binned to 20 km/s (similar to the velocity dispersion of 20.2 km/s reported in M08 for W32604).\raggedright
\label{table1}
\end{table*}


\section{Discussion}
\label{sec:discussion}

From Table \ref{table1} we can see that our non-detections of CO in the ALMA spectra of these clusters can be translated into upper limits on the H$_2$ mass estimates which are of the order of $\sim9\%$ of the stellar mass. Having said that, we note that this estimate is subject to the choice of $ X_\text{CO} $. For this study we have used the value reported for the Milky Way disk by \cite{Bolatto:2013p2560}. However, several studies have shown that the $ X_\text{CO} $ in the Antennae is lower by a factor of 2--10 (e.g. \citealt{Zhu:2003p2569,Bisbas:2014p2567}) than the Milky Way value. As a consequence the \hdos~mass estimate will decrease by the same factor if $ X_\text{CO} $ decreases, so \emph{the values of molecular gas reported in Table \ref{table1} are strict upper limits}.


As mentioned in \S\ref{sec:intro}, given the age range of these clusters (50--200 Myr) we are only able to place constraints on the AGB scenario. However, this sample spans the critical timeframe where the second generation of stars is expected to be formed for this choice of \emph{polluters} (30--300 Myr, see  \S\ref{sec:intro}). There are several variations of this scenario that adopt AGB stars as the source of enrichment. In the following subsection we outline the essentials of an ``AGB scenario" and some distinctions between different scenarios.

\subsection{AGB scenario: The essentials}
\label{sec:agb}

\begin{enumerate}
\item A first generation (1G) is born as a simple stellar population, i.e. all the stars with the same age, and homogeneous (\emph{pristine}) chemical composition. 
\item The remaining pristine gas that formed the 1G, is expelled from the cluster at some time during the following $\sim30$ Myr by the winds of massive stars and core-collapse SNe (R08).
\item A second generation (2G) starts to form. The first stars to be born will be from the pure (i.e. undiluted) ejecta of the most massive $(\sim 8-9$ \msun) AGB stars\footnote{a.k.a. super-AGB stars}. This material will be forming the 2G stars with the most extreme abundance patterns, that is, the most He-rich population which is also the one with the most O-poor stars observed today in old GCs. We will refer to this population as the \emph{extreme} 2G (the extreme stars identified by \citealt{Carretta:2009p2165}).\\
Curiously, the 2G stars are assumed to span a rather narrow mass range
$0.1-0.8$ \msun~(D08). This preference for a truncated initial mass function (IMF)  for the 2G stars in these models, would have two main consequences 1) this generation of stars will not have type II SNe that would clear the gas, interrupting the formation of the 2G; and 2) minimise the ``mass budget problem", i.e. the fact that one would require a 1G many times more massive than the amount of stars with primordial abundances seen today. Both consequences would be treated again in (vi) and (vii) respectively.
\item Only after the extreme 2G is born, the cluster starts accreting pristine material from a surrounding reservoir. The second star formation episode will continue forming stars, this time with 
\emph{intermediate} chemical abundances (i.e. abundances between the extreme and pristine ones) from the ejecta of the lower mass $(\lesssim 8$ \msun) AGB diluted with this accreted material.\footnote{CS11 in their model predicts a continuum accretion of pristine gas just after the 1G is formed, but the star formation is inhibited during the first 100--200 Myr given that the Lyman-Werner 1G photons have been keeping the gas warm. Due to this (ii) and (iii) never took place in this variation of the AGB scenario. \cite{DErcole:2011p2572} argue that this mechanism (continuous accretion) would not be able to reproduce the \emph{extreme} populations observed in GCs, see \S\ref{sec:accretion}.}\\
We need to underline this point, given that the dilution is required in all the scenarios that share AGB  stars as polluters. \emph{Although several attempts have been presented, to date it still remains quite equivocal where was this pristine material stored, what triggers the accretion, what process regulates its rate, and what mechanism ends it.} However, it is possible to constrain the nature of this sequence of events with the current knowledge of the field, for example:
\begin{itemize}
\item Where does this pristine gas reservoir come from? This element is not clear in the scenarios so far. D08 implies that the 1G cluster is surrounded by some pristine gas, and the accretion starts after the extreme 2G is formed. On the other hand CS11, argue that the young GC continuously accretes material, as it sweeps the surrounding  interstellar medium (ISM) in a competing process against ram pressure. Later, after the number of energetic photons from the 1G drops, the gas that has been accumulated in the cluster's potential is allowed to cool down to form stars (CS11).
\item In \cite{DErcole:2011p2572} (hereafter D11), the authors conclude that the maximum amount of pristine gas that contributes to the 2G formation is at most equal to 10\% of the 1G mass in order to reproduce the extent of the observed abundance patterns. This limit was obtained assuming high star formation efficiencies ($\sim100\%$), if lower values were in play in the 2G formation, the mass of the gas reservoir should increase (further discussed in \S \ref{sec:toy_model}).
\item The expulsion of the pristine gas mentioned in step (ii) by core-colapse SNe, should have been through a very distinct mechanism in the sense that while it is able to expel this pristine material (now polluted with Fe-peak and $\alpha$-elements) away from the cluster, it does it in such way that this polluted material never reaches the pristine gas reservoir. This has been explored by R08 and \cite{Renzini:2013p2586}, concluding that if only 0.2\% of the core-collapse SN ejecta would have reached the pristine gas reservoir the 2G of the cluster would exhibit [Fe/H] inhomogeneities like the ones observed in $\omega$Cen, M22 and Terzan 5.\\
It has been put forward that some asymmetries in the gas distribution could prevent the Fe-polluted material to reach the reservoir, avoiding the contamination (cf. D08). In this case the core-collapse SNe leave an ``hour glass" cavity after clearing the leftover gas from the 1G, allowing the uncontaminated material around the ``hour glass" to collapse back into the cluster for step (iv).
\item The accretion must be an accelerated process. The theoretical yields of the AGB stellar models predict a correlation between Na and O, in their ejecta, as a function of stellar mass/time. Given this, after the \emph{extreme} 2G is formed, the pristine gas accretion rate should increase rapidly in time, in order to be able to dilute the material not just compensating their natural correlation trend but to a higher degree, in order to reproduce the observed Na-O anti-correlation.
\end{itemize}
\item It is assumed that the star formation of the 2G took place in bursty episodes. This premise in the models is backed by the evidence of discrete (i.e. non continuous) He abundances in the subpopulations of some massive GCs (e.g. NGC 6397, 2808 \& 6752 - \citealt{Milone:2012p2160,Milone:2012p2735,Milone:2013p2709}). Furthermore, this assumption also agrees with the recent discovery of a few massive GCs (e.g. NGC 6752 \& 2808 cf. \citealt{Carretta:2012p2708,Carretta:2014p2707}) showing clear discreteness on their light element abundance patterns, specifically, the Mg-Al anti-correlation.
\item The formation of the 2G should end before the low-mass $(\lesssim 3$ \msun) AGB stars produce large amounts of C, which is especially the case for low metallicity AGBs (cf. R08, \citealt{Dercole:2010p2571}). The production of C, would affect the sum of CNO elements which appears to be constant (within a factor of $\sim1.5$) among primordial and enriched stars in GCs (e.g. \citealt{Carretta:2005p2737}). This requieres a mechanism which should be synchronised to start operating just before these lower mass AGB go into action, otherwise (in principle) the star formation would continue. Regarding this, most authors have agreed on the role of prompt ($\sim 10^8$ yr) Ia SNe on this matter.\\
In CS11, the 2G star formation happens suddenly. Here an instantaneous burst of star formation gives birth to the 2G of stars after the number of Lyman-Werner photons (912\AA~$< \lambda <$ 1100\AA) drops sufficiently to allow the star formation (200--300 Myr after the birth of the 1G, CS11). The star formation ends when the type II SNe from the 2G and prompt Ia SNe from the 1G start to go off, keeping the cluster gas free for the rest of its life. In a similar manner, for R08 the formation of the 2G is likely span up to 300 Myr as well. Constrastingly, in D08 the prompt Ia SNe are expected to end the 2G formation episodes $\sim80$ Myr after the 1G star formation event. In this model there will be no type II SNe from the 2G IMF since all the stars have masses well bellow the limit for core-collapse SNe (D08). 
\item After the 2G is formed in these models, the young cluster ends up with just a small fraction of the stars showing the chemical signature of the AGB material. In order to match the observed high fraction (close to 1:1) of enriched/non-enriched stars (i.e. ratio of ``first-to-second generations of stars") these scenarios assume that GCs underwent strong mass-loss and lost most of their first generation stars, usually 90\% or more, which means that GCs were singificantly more massive at birth than seen today. Furthermore, these models usually assume that the 2G stars had a truncated mass function.\\
In D08, the authors claim that the cluster should have been about 10 times more massive at birth, however, this value has some severe underlying assumptions. One can see from Fig. 4 of D08, that the stellar mass of the 2G is just $\sim10^{4.5}$ \msun, this corresponds to a $\sim1/30$th of their 1G mass ($10^6$ \msun). So only if the current (old) cluster has a particularly low fraction of enriched/non-enriched stars (1:3) and only if 1G stars were lost (all 2G stars were kept in the cluster) one can reach this value for the mass of the 1G (originally 10 times more massive). Alternatively, if the current cluster has an even fraction (1:1) of enriched/non-enriched stars the cluster should have been 30 times more massive at birth (again, this value assumes no 2G stars escaped the cluster during its dynamical evolution).\\
Additionally, as mentioned in (iii), D08 is assumed a truncated IMF (only forming stars with $0.1-0.8$ \msun) for the 2G. If the 2G would have formed following a conventional (i.e. \citealt{Kroupa:1993p2738}) non-truncated IMF (i.e. stars form with masses between 0.1--100 \msun) the mass of the first generation should scale by a factor of 2 (i.e. 1G should have been 20 or 60 times more massive in the past depending of the assumed fraction of enriched/non-enriched stars of 1:3 and 1:1 respectively).\\
On the other hand, in CS11 the young GCs had a similar masses than the ones observed today (D11), while \cite{Conroy:2012p2577} support that the GCs were at least 10--20 times more massive at birth.\\
Although there has not been reported any evidence of a strongly non-standard IMF in GCs or YMCs (cf. \citealt{Bastian:2010p2000}) 
 this option is difficult to exclude. However, in the last few years there have been some constraints reported regarding the strong mass-loss.
\end{enumerate}

\cite{Larsen:2012p2565} and \cite{Larsen:2014p2566} studied three dwarf galaxies, and looked at the amount of GCs stars which could have been lost and now form part of the field populations of their host galaxies. In both studies the authors found high GCs-to-field stars ratios, concluding that these GCs could only have been at most 4--5 times more massive initially.
 These values are upper limits, given that it was assumed that all of the field stars that share the same metallicity as the GCs (in a broad metallicity range), formed in these clusters (i.e. no field stars formed with similar abundances) nor did they take into account that lower mass clusters would likely have been formed at the same time and have since disrupted (e.g. \citealt{Kruijssen2015}), contributing to the field population.
Furthermore, the recent model by \cite{Kruijssen2015}, putting GC formation into a hierarchical galaxy assembly context, also found that clusters that survive to the present epoch were likely to be only 2--3 times more massive than currently observed.

In the following subsection we will present a ``toy model" designed to explore the viability and consequences of the AGB scenario, given the constraints we have just presented regarding the stellar mass and upper limits on any potential gas within the observed clusters.

\subsection{Toy model}
\label{sec:toy_model}

The ``toy model" is based on the most massive young globular cluster in our sample, W32187 with a mass of  $\sim2\times10^6$ \msun. This cluster lies within the age range where a second episode of star formation is expected to occur in the AGB model. We adopt the initial mass upper limits discussed above \citep{Larsen:2012p2565,Larsen:2014p2566,Kruijssen2015}, and assume that the current mass of the cluster is $\sim4$ times more massive than it will be in $\sim10$ Gyr\footnote{This value differs the original D08 scenario (as seen in \S\ref{sec:agb}), however, with this adjustment we take into account the \cite{Larsen:2012p2565,Larsen:2014p2566,Kruijssen2015} results.}, i.e. assuming that the current cluster stellar mass is the initial mass. This means that as an old GC, this cluster will have a mass of M$_\star=5\times10^5$ \msun.
If we also assume a 1:1 ratio for the number of first-to-second generation stars at an age of $\sim10$~Gyr, this would correspond to a mass of the second generation (2G) stars of M$_\star^{\text{2G}}=2.5\times10^5$ \msun. This is a lower limit to the amount of gas that would need to be present, integrated over time. However, if we adopt a standard star-formation efficiency for this cluster of SFE=$1/3$ (e.g. \citealt{Lada:2003p2573}), this means that M$_{gas}^{\text{2G}}=7.5\times10^5$ \msun~of molecular gas is required to give birth to the 2G. The rest of this gas i.e. the other 2/3 that will not end up in stars, is expelled from the cluster (see \S\ref{sec:hmxb}) after the star formation takes place.
Given these assumptions, \emph{how much molecular gas would we expect at any given time within this cluster?}

To answer this question we consider the window where star formation can take place according to D08, $\Delta t=50$ Myr (i.e. from 30--80 Myr, see \S\ref{sec:agb}). If the star formation is extremely rapid, in the sense that any gaseous material within the cluster is used in the star formation very quickly, e.g. $\Delta t^{\text{2G}}_{SF}=1$ Myr\footnote{We note that if we assume a slighter shorter timescale of $\Delta t^{\text{2G}}_{SF}=0.6$ Myr our simple model is in good agreement with D08 predictions.}, we would expect
to find M$_{gas}^{\text{2G}}=1.5\times10^4$ \msun\ (i.e., $1/50^{\rm th}$ of the total gas mass) at any given moment during this $50$~Myr time window, which lies below our detections limits of $\sim1\times10^5$ \msun. However, if the star formation event lasts longer, i.e. that any gas within the cluster is present for say $\Delta t^{\text{2G}}_{SF}=10$ Myr, before being consumed in star-formation, the amount of gas at any given time would increase to M$_{gas}^{\text{2G}}=1.5\times10^5$ \msun, which is above our (conservative/overestimated) detection limit for this cluster of $1.38\times10^5$ \msun. Moreover, if such clusters have undergone a significant mass loss (as claimed in the D08 model), then we would not be looking at the initial mass, but rather the present day values (contrary to what was assumed here). Hence, if the initial mass was significantly higher, we would expect more gas to be present in the cluster, for example if the cluster has already lost 50\% of its initial mass, then these estimates would increase by a factor of 2.

CS11 suggest that $\Delta t^{\text{2G}}_{SF}$ must be long ($\sim100-300$~Myr) as the Lyman-Werner flux density from the first generation stars will be too high to allow the gas/dust to cool sufficiently to form stars.  In this case, we would have expected to detect large amounts of gas ($>10\%$ of the stellar mass) within all three of the clusters observed in the present work, in contrast to what is observed (an upper limit of $<9\%$ of the stellar mass).

This simple ``toy model" demonstrates the general bounds that may be expected in the AGB scenario.  The actual amount of gas within the cluster depends critically on 1) the assumed SFE and 2) the time of which the ISM may exist within the cluster before being expelled or used to form 2G stars.

\subsubsection{Summary of the toy model and comparison with previous works}

Our ``toy model" was developed to find the \emph{minimum amount of gas} during the formation of a 2G needed to form the number of stars with enriched abundances (a.k.a. 2G stars in the AGB scenario) seen today in a GC. In this model we have assumed that at every $\Delta t^{\text{2G}}_{SF}$, 1/3 of all gas (regardless of its origin i.e., either ejected by AGB stars or accreted from the pristine gas reservoir) is turned into stars, while the remaining 2/3 of the gas will be lost from the cluster (this could be by means of any of the mechanisms mentioned in \S\ref{sec:hmxb}). Nevertheless, this is just an assumption. However, if this material is not cleared away from the cluster, 
but instead remains within the cluster, it could build up over time
.

This alternative has been modelled and thoroughly developed by D08. Here, the authors assume that the gas that has not found its way into a star (i.e. 2/3 of the total gas in our case) remains in the cluster. Perhaps this material is too warm to form stars yet, but the key issue here is that it has not been expelled from the cluster (in D08 this gas will eventually cool and start forming new stars)\footnote{If the gas can cool, and remains in the cluster, it will eventually form new stars. Hence, even if the SFE is low per unit time in this model (D08), it can approach unity over the full formation epoch of the 2G (i.e. assumed to be 50 Myr in this case).}. This process results in the accumulation of gas within the centre of the cluster. However, since the gas mass and density are increasing, D08 assume that this will also have an effect on the star formation rate (SFR), with higher densities resulting in higher SFRs. This treatment is physically motivated, but of course adds some additional assumptions and parameterisations that are beyond the scope of our simple toy model.

However, if some kind of feedback is considered (which is not the case in D08 - see \S\ref{sec:hmxb}) and would have kept the gas from cooling (so that it cannot form new stars, but it does remain in the cluster), then the gas mass will increase as a function of time inside the cluster (e.g., as in the CS11 scenario).
 The longer the gas accumulates within the cluster the easier it would be to detect it. Additionally, if the gas temperature increases (due to the feedback mechanism, e.g., Lyman-Werner flux) then the gas would be easier to detect, for a given amount of gas mass, with the type of observations used in the current work. 
 Although perhaps our toy model might be considered less physically motivated than assuming some fraction of gas is retained, \emph{we note that this approach provides a lower limit to the amount of gas expected in the cluster at any given time.}
 
Finally, we must remark that our unsophisticated model is in no way meant to replace the ones provided by D08/\cite{Dercole:2010p2571} or CS11 which are much more physically motivated, which is the reason we why we have chosen to compare against them. Nevertheless, we have shown that this simple description in our ``toy model'' is able to reproduce the same processes as the more elaborate model of D08, without the need to include further unconstrained parameters. Here we have shown that a good agreement is achieved between D08 model and our ``toy model", if one adopts a relatively short $\Delta t^{\text{2G}}_{SF}$ of 0.6 Myr.




\subsection{Accretion of pristine gas}
\label{sec:accretion}

CS11 have presented a model for how pristine gas may be continuously accreted at a (nearly) constant rate by the GC in order to dilute the AGB ejecta within the cluster, and form 2G stars (such strong dilution is required on chemical grounds, e.g. \citealt{Dercole:2010p2571}). In the CS11 model, young GCs contain 10\% of their initial stellar mass in gas within the cluster, and this gas acts as a net as the cluster passes through the galactic ISM, picking up further pristine material.  In this case, the clusters would always be expected to have $>10\%$ of their stellar mass in gas within them, which would be above our detection limit. Hence, this scenario is not consistent with the observations presented here. 

Moreover, \cite{DErcole:2011p2572} have shown that any model with nearly constant accretion of pristine gas can not reproduce the observed abundance trends found in GCs. Instead, the accretion of pristine gas must be finely tuned, i.e. it cannot contribute significantly until $\sim20$~Myr after the 2G begins to form.  This allows for the most enriched 2G stars to form entirely from (super)AGB ejecta, while subsequent 2G stars form from a combination of AGB ejecta and pristine gas, which dilutes the abundances, and turns the predicted correlation between Na-O from AGB stars, into an anti-correlation (as mentioned in \S\ref{sec:agb})



\subsection{D08 mass prediction}

From Fig. 3 of D08, we obtained an estimate of the amount of gas expected to be found in a young (100 Myr) massive ($10^6$ \msun) cluster, integrating the gas density profile in a radius of 25 pc (approximately the size of our beam readius). We obtained $\sim1500$ \msun~of gas, which corresponds to a reservoir of molecular gas of $\sim0.15\%$ of the stellar mass at an age of $100$~Myr. This implies an extremely rapid $\Delta t^{\text{2G}}_{SF}$ ($\Delta t^{\text{2G}}_{SF}\simeq 6\times10^5$yr, see \S\ref{sec:toy_model}).  As discussed in CS11, such rapid turn-over from accreted/ejected gas to stars, is likely optimistic, as it ignores the large heating effect of evolving first generation stars, i.e. stellar feedback.\\
In D08 the only heating source in the standard model was the thermalisation of the kinetic energy of the stellar winds. D08 ran additional models with ``extra energy sources" (e.g. X-ray binaries and planetary nebulae nuclei) and found that such heating may terminate any 2G star formation (above certain threshold). However, detailed models are required to understand the exact efficiency of stellar feedback on any material residing in the cluster. Such models have yet to be done for young GCs, however \cite{McDonald:2015p2751} have run these models for older GCs and have found that stellar feedback is very efficient at clearing out and gas/dust within the cluster. Since the feedback at young ages ($<200$ Myr) is significantly higher than at $\sim10$ Gyr, it is likely that it can provide an effective gas/dust clearing mechanism, potentially explaining the results presented here.
\\
Nonetheless, due to the lack of heating included in the current D08 model, the accreted/ejected gas can sink in the very centre of the cluster ($<0.1$ pc), where the gas density becomes extremely high, leading to rapid and efficient star-formation. Additionally, this model assumes that the 2G of stars only make up $\sim3$\% of the initial mass of the first generation, which needs to be increased in order to match the relative numbers of 1G/2G stars in some clusters.

\subsection{Why are YMCs gas free?}
\label{sec:hmxb}

As mentioned in the previous section, D08 studied the possibility of models with extra energy sources (only the wind thermalisation energy was included in the standard model). They concluded that if a critical value of $Q_{cr}=2\times10^{36}$ erg/s was exceeded, a $10^6$ \msun~cluster would abruptly transition from a model hosting a cooling flow, which would be collecting the AGB ejecta in the centre of the cluster, to a model hosting a wind (i.e. expelling the material from the cluster). Furthermore, they arrive at the conclusion that if a luminosity of $Q=1.38\times10^{37}$ erg/s would to be reached within a massive ($10^7$ \msun) cluster, the immediate onset of a fast wind would preclude the formation of any 2G\footnote{The equivalent value for a $10^6$ \msun~cluster was not provided in D08, but we assume it would be below this $Q=1.38\times10^{37}$ erg/s limit.}. These authors proposed X-ray binaries as a possible source of this energy.\\
\cite{Power:2009p2770} studied the role of high mass X-ray binaries (hereafter, HMXBs) in the first few hundred Myr of GCs. In their Fig. 2, one can see the evolution of the H ionizing photon rate over time, produced by the HMXB population of a $10^6$ \msun~cluster. In this plot one can see that even if only 
50\% of the high mass binaries that survive the SNe of the primary star of the binary system become HMXBs (it is assumed that only 1/3 of binaries will survive the primary SNe, i.e. $\sim15\%$ of the initial high mass binary population), 
one would expect to have $Q\sim1.5\times10^{40}$ erg/s and $Q\sim2.2\times10^{38}$ erg/s at 30 and 80 Myr respectively. From this we see that, during the timescale where the 2G is expected to be form in D08, these $Q$ values are several order of magnitude above the threshold ($Q_{cr}\sim10^{36}$ erg/s) necessary to stop the star formation of the 2G in these $10^6$ \msun~clusters. However, a crucial parameter that requires further modelling is what fraction of this energy is transferred to the ISM within the cluster, and what fraction escapes the cluster directly.\\
This makes HMXBs a potential candidate to explain why these clusters appear to be gas free. This possibility is reassured when considering that in these environments a single HMXB source has a typical luminosity in excess of $L_X\sim10^{38}$ erg/s (e.g. \citealt{Fabbiano:2001p2772,Wolter:2004p2773,Rangelov:2012p2771}).\\
Other plausible mechanism able to remove the gas from the cluster could be ram pressure striping by the ISM around the cluster, as shown in CS11. Originally, this mechanism would not be effective for the D08 scenario as it is, due to the fact that in this model most of the gas gathered by the cooling flow would be concentrated in the innermost region ($\lesssim0.1$ pc) of the cluster. Such an extremely concentrated volume of gas would not be sensitive to the influence of ram pressure, since this mechanism is heavily dependant on the surface area/cross section of the gas within the cluster. Hence this effect can be neglected for the current version of this model. Notwithstanding, this changes if the gas were to occupy a larger volume within the cluster. For example, if this gas were to be exposed by some heating (some sort of stellar feedback, currently not considered in this model e.g., Lyman-Werner photons) it would be reasonable to expect the density of this gas to decrease, and as a consequence, covering a larger volume within the young cluster. Therefore, if this (heated)gas would now cover the central 1 pc of the cluster, the new surface of this gas would have increased by a factor of 100, and the effect of the ram pressure would increase as well in the same proportion. To summarise, if the gas ejected by the polluters stars were not to be concentrated in a very small volume at the centre of the cluster, it is likely that the ram pressure striping by the ISM surrounding the cluster would play a non-negligible role in the expulsion of some the gas out of the cluster.\\
For the moment, we can not provide a definitive answer to why are YMCs gas free, and in order to reach a definite conclusion on what could be the possible explanation of this phenomenon, we require further studies that are beyond the scope of the current paper. Nevertheless, we have shown that at least these two possibilities could play a role in its answer and should be contemplated in future studies.

\section{Conclusions}
\label{sec:conclusions}

We use ALMA CO(3--2) observations to search for cool gas within three young massive clusters in the Antennae merging galaxies. Some scenarios for the formation of the observed multiple populations in GCs predict that such YMCs should contain significant amounts of cool gas ($\ge10$\% of the initial stellar mass of the system) within them in order to form multiple generations of stars (e.g. D08, CS11).  The observations of clusters W31730, W32604 and  W32187 used in the current work provide upper limits of $\sim1\times10^5$ M$_\odot$ of H$_2$ gas within the three clusters or 7.4, 8.8, and 7.3 \% of their stellar mass, respectively.  We note, however, that the adopted $X_\text{CO}$ may be overestimated by up to an order of magnitude, which would decrease the measured upper limits by the same amount.

We have estimated the amount of gas that is expected to be present within the clusters, in various versions of the AGB scenario, and found that the exact amount depends crucially on a number of assumptions and uncertainties (e.g. the amount of first generation stars lost, star-formation efficiencies, stellar IMF variations), hence we cannot reach definitive conclusions regarding the AGB model in general. 

For the D08 scenario, the results from W31730 and W32604 clusters are consistent with their prediction. However, if one were to adopt less extreme values for the main parameters in play in the model (e.g. IMF, SFE, fraction of enriched/non-enriched stars, 1G and 2G mass loss, etc.) instead of the adopted by the authors, one would expect significant amounts of gas in these young clusters (as shown in \S\ref{sec:discussion}) and would be easily detectable by our observations.
On the other hand, cluster W32187 (200 Myr) in principle would lie outside the timescale for the formation of the 2G adopted in this scenario (30--80 Myr) established by the first SNe Ia explosions. Having said that, we should emphasise that the timescale of these prompt Ia SNe is highly unconstrained, and in this model there is nothing inherent in the 80--100 Myr values adopted. Current evidence suggests, that these events could start as soon as just $\sim40-50$ Myr or as late as 400 Myr after the 1G was formed (e.g. \citealt{Ruiter:2013p2769,Maoz:2010p2776,Brandt:2010p2775}). Hence, the results from this particular cluster have a constraining affect in D08 as well. Ultimately, the CS11 AGB scenario, predicts gas masses well in excess of the upper limits provided here (during these ages), hence our results are in strong tension with this model.

The lack of evidence of any cold gas within these clusters could be explained in two ways 1) The actual levels are below our detection limits or 2) that the clusters are very efficient at removing the gas. This could be due to heating by stellar feedback and/or subsequent stripping i.e. ram pressure (CS11, \citealt{Power:2009p2770,McDonald:2015p2751}). However, further full radiative transfer calculations are required to assess the precise magnitude of these effects.

Although the constraints placed by our ALMA observations are not conclusive in ruling out the AGB scenario, we note that they are in agreement with a growing number of studies that have not found evidence of extended (or multiple) star-formation episodes within massive clusters.  For example, no evidence of ongoing star formation has been detected (down to limits of few percent of the mass of the first generation) in young ($10^6-10^9$ yr) massive ($10^4-10^8$ \msun) clusters (cf. \citealt{Bastian:2013p2022}). Also, \citet{CabreraZiri:2014p2558} used an integrated optical spectrum of a massive ($>10^7$ \msun) young ($\sim100$ Myr) cluster, in order to  estimate its star-formation history.  The cluster was found to be consistent with single stellar generation, with no evidence of a secondary burst down to mass ratios of 10--20\% of the current cluster mass.  Additionally, \citet{Bastian:2013p2199} and \cite{Niederhofer2014} have searched for evidence of age spreads in resolved massive LMC/SMC clusters, and found that all are consistent with a single burst of star-formation.  Altogether, the evidence from observations of YMCs places strong constraints on the AGB scenario, ruling out large areas of parameter space.


\section{Acknowledgements}

We would like to thank C. Power for helpful discussions. This paper makes use of ALMA data: ADS/JAO.ALMA\#2011.0.00876.S. ALMA is a partnership of ESO (representing its member states), NSF (USA) and NINS (Japan), together with NRC (Canada) and NSC and ASIAA (Taiwan), in cooperation with the Republic of Chile. The Joint ALMA Observatory is operated by ESO, AUI/NRAO and NAOJ. The National Radio Astronomy Observatory is a facility of the National Science Foundation operated under cooperative agreement by Associated Universities, Inc. NB is partially funded by a Royal Society University Research Fellowship. 

\bibliographystyle{mn2e}
\bibliography{cabrera-ziri15}

\end{document}